\documentclass[conference]{IEEEtran}
\IEEEoverridecommandlockouts
\usepackage{cite}
\usepackage{amsmath,amssymb,amsfonts}
\usepackage{algorithmic}
\usepackage{graphicx}
\usepackage{textcomp}
\usepackage{xcolor}
\def\BibTeX{{\rm B\kern-.05em{\sc i\kern-.025em b}\kern-.08em
    T\kern-.1667em\lower.7ex\hbox{E}\kern-.125emX}}

\usepackage{lineno}
\usepackage{amsthm}
\usepackage{amssymb}
\usepackage{amsmath} 
\usepackage{bm}      
\usepackage{booktabs}
\usepackage{scalerel}
\usepackage{multirow}
\usepackage[ruled,longend]{algorithm2e}
\usepackage{threeparttable}
\usepackage{comment}

\usepackage{amsthm}
\usepackage{lipsum} 
\usepackage{xcolor}

\usepackage{hyperref}
\hypersetup{
    colorlinks=true,
    citecolor=black,
    linkcolor=black,
    filecolor=black,
    urlcolor=black
    }

\newtheorem{theorem}{Theorem}[section]

\newtheorem{Definition}{Definition}[section]

\usepackage{epstopdf}

\begin{document}
\begin{sloppypar}


\title{Form-Finding and Physical Property Predictions of Tensegrity Structures Using Deep Neural Networks \thanks{The research of J.Q. is supported by NSF grant DMS-1941197.}}



\author{\IEEEauthorblockN{1\textsuperscript{st} Muhao Chen}
\IEEEauthorblockA{\textit{Department of Mechanical and Aerospace Engineering} \\
\textit{University of Kentucky}\\
Lexington, KY 40506\\
muhaochen@uky.edu}
\and
\IEEEauthorblockN{2\textsuperscript{nd} Jing Qin}
\IEEEauthorblockA{\textit{Department of Mathematics} \\
\textit{University of Kentucky}\\
Lexington, KY 40506\\
jing.qin@uky.edu}}

\maketitle

\begin{abstract}
In the design of tensegrity structures, traditional form-finding methods utilize kinematic and static approaches to identify geometric configurations that achieve equilibrium. However, these methods often fall short when applied to actual physical models due to imperfections in the manufacturing of structural elements, assembly errors, and material non-linearities. In this work, we develop a deep neural network (DNN) approach to predict the geometric configurations and physical properties—such as nodal coordinates, member forces, and natural frequencies—of any tensegrity structures in equilibrium states. First, we outline the analytical governing equations for tensegrity structures, covering statics involving nodal coordinates and member forces, as well as modal information. Next, we propose a data-driven framework for training an appropriate DNN model capable of simultaneously predicting tensegrity forms and physical properties, thereby circumventing the need to solve equilibrium equations. For validation, we analyze three tensegrity structures, including a tensegrity D-bar, prism, and lander, demonstrating that our approach can identify approximation systems with relatively very small output errors. This technique is applicable to a wide range of tensegrity structures, particularly in real-world construction, and can be extended to address additional challenges in identifying structural physics information.
\end{abstract}


\section{Introduction}\label{sec:intro}

The concept of tensegrity is defined by a stable arrangement of compressive (bars or struts) and tensile (strings or cables) elements \cite{skelton2009tensegrity}. Extensive research into tensegrity has revealed several advantages, including its lightweight nature \cite{chen2023minimal}, adjustable stiffness \cite{habibi2023effects}, and significant morphing capabilities \cite{song2024dynamic}. These attributes have enabled its use in diverse applications such as airfoil design \cite{shen2023markov}, metamaterials development \cite{amendola2019mechanical}, fish robot creation \cite{chen2019swimming}, rover innovation \cite{zappetti2022dual}, tower engineering \cite{chen2023analysis}, climbing robot development \cite{kobayashi2022soft}, robotic manipulator \cite{woods2023design}, and self-assembling modular robots \cite{zhao2023starblocks}. Consequently, tensegrity has proven to be a versatile foundation for designing novel, lightweight, and deployable structural systems.

Despite the increasing popularity of tensegrity structures, significant challenges persist in determining their geometrical configurations through a process known as form-finding. Early pioneers in tensegrity, i.e., Snelson \cite{snelson1965continuous}, primarily used regular, convex polyhedra as the basis for exploring new configurations of tensegrity structures. However, physical models indicate that the shape of a tensegrity structure based on specific polyhedra, such as the truncated tetrahedron or the expandable octahedron, deviates from the original polyhedral form \cite{tibert2011review}. Consequently, proper form-finding methods are crucial for accurately determining the equilibrium configurations of even basic tensegrity structures \cite{motro1992tensegrity}. Form-finding methods for tensegrity structures have been extensively studied.

Most of the work begins by developing analytical approaches or efficient algorithms to tackle their nonlinear statics. Examples include the force density method \cite{wang2021form}, dynamic relaxation method \cite{he2024modified}, Levenberg-Marquardt method \cite{yuan2017form}, and Laplacian matrix based method \cite{dong2019inverse}. However, these methods often encounter limitations when applied to actual physical models. Such limitations typically stem from various practical issues, such as imperfections in the manufacturing of structural elements, assembly errors, and the effects of material fatigue, erosion, and creeping. These factors, along with other unpredictable reactions, are difficult to model analytically, leading to significant discrepancies in the calculated physical properties of real tensegrity structures.

Recently, machine learning has gained significant traction due to its remarkable ability to process large-scale datasets. Several initiatives have been undertaken to apply machine learning to the problem of form-finding in tensegrity structures. For instance, Lee et al. substituted the traditional force density method with deep neural networks that predict element lengths from force densities \cite{lee2022deep}. Zalyaev et al. developed a machine learning pipeline that includes feature extraction and regression for Tensegrity Form Finding \cite{zalyaev2020machine}. Additionally, Kim et al. employed a Monte Carlo-based learning algorithm to control the locomotion of tensegrity robots through form-finding \cite{kim2015robust}. However, very few of them could provide any physical properties of the tensegrity structure other than nodal coordinates. In many applications, physical properties such as tension forces and frequencies are key considerations in tensegrity structure design to ensure structural stability, load distribution, and overall performance. In addition, when dealing with a large number of tensegrity structures within the same category, data-driven approaches, such as DNNs, would be more efficient in simultaneously inferring tensegrity forms and physical properties based on the cable length changes than model-driven methods, such as solving static equilibrium equations. Motivated by these insights, we aim to develop a novel DNN-based modeling framework for predicting aspects of tensegrity form-finding and physical properties, including nodal coordinates, tension forces, and frequencies.

The rest of the paper is structured as follows: Section~\ref{tenseg} introduces the notations and governing equations of tensegrity structures. Section~\ref{dnn} details the DNN model, including the testing and prediction framework. Section~\ref{exp} examines three case studies involving a tensegrity D-bar, prism, and lander and discusses the results. The paper concludes with Section~\ref{conc}, which offers a discussion of the findings and outlines directions for future research.

\section{Governing Equations of Tensegrity Structures}\label{tenseg}

Before exploring the physics of tensegrity structures, such as equilibrium states affected by changes in string force, forces in all the structural members, and natural frequencies, we introduce the notations and governing equations relevant to any tensegrity systems.

\subsection{Tensegrity Structure Notations}
\label{Generalized}

\begin{Definition}[Nodal Coordinates]
The nodes in a tensegrity structure can be represented in both vector and matrix forms: the vector form is denoted as:
$\bm{n} = \begin{bmatrix} \bm{n}_1^\mathrm{T} &\bm{n}_2^\mathrm{T} & \cdots & \bm{n}_{n_n}^\mathrm{T}\end{bmatrix}^\mathrm{T}\in \mathbb{R}^{3n_n}$, and the matrix form is $\bm{N} = \begin{bmatrix} \bm{n}_1 & \bm{n}_2 & \cdots & \bm{n}_{n_n} \end{bmatrix} \in \mathbb{R}^{3 \times n_n}$. Here, $n_n$ represents the total number of nodes, and
$\bm{n}_i = \begin{bmatrix} x_i & y_i & z_i \end{bmatrix}^\mathrm{T}$ corresponds to the $ x$-, $y$-, and $z$-coordinates of the $i$-{th} node ($ i = 1, 2, \ldots, n_n $).
\end{Definition}

\begin{Definition}[Connectivity Matrices]
Connectivity matrices represent the patterns of connections between bars and strings in the nodes of a structure. The matrices for bars and strings are denoted as $\bm{C}_b \in \mathbb{R}^{n_b \times n_n}$ and $\bm{C}_s \in \mathbb{R}^{n_s \times n_n}$, where $n_b$ and $n_s$ indicate the total numbers of bars and strings, respectively. For each structure element $k$ connecting nodes $\bm{n}_i$ and $\bm{n}_j$ (where $i < j$), the connectivity matrices include a ``$-1$'' at the $k$-{th} row and $i$-{th} column, a ``$+1$'' at the $k$-th row and $j$-th column, and zeros elsewhere in the same row. The indices $i$ and $j$ range from 1 to $n_n$, and $k$ ranges from 1 to $n_b$ for bars and 1 to $n_s$ for strings. The complete connectivity matrix $\bm{C} \in \mathbb{R}^{n_e \times n_n}$ (with $n_e = n_s + n_b$ representing the total number of structural elements) is formed as $\bm{C} = \begin{bmatrix} \bm{C}_b^\mathrm{T} & \bm{C}_s^\mathrm{T}\end{bmatrix}^\mathrm{T}$.
\end{Definition}

\subsection{Tensegrity Governing Equations}

\begin{theorem}[Tensegrity Statics]\label{theorem_dyn}
The static equilibrium equations for a tensegrity structure, expressed in terms of the nodal vector $\bm{n}$ and the member force vector $\bm{t}$, are given by:
\begin{align}\label{tts_equili}
\bm{E}_{a}^{T}\bm{K}\bm{n} & =\bm{E}_{a}^{T}(\bm{f}_{e x}-\bm{g}),\\\label{tts_equilit}
\bm{E}_{a}^{T}\bm{A}_{t} \bm{t} & =\bm{E}_{a}^{T}(\bm{f}_{e x}-\bm{g}),
\end{align}
where $\bm{K} =(\bm{C}^T\hat{\bm{x}}\bm{C})\otimes\bm{{ I}}_3$ and $\bm{A}_{t}=(\bm{C}^{T} \otimes \bm{I}_{3}) \bm{b.d.}(\bm{NC^T})\hat{\bm{l}}^{-1}$ represent the stiffness matrix and equilibrium matrix. Here, $\otimes$ denotes the Kronecker product, and the operator $b.d. (\bm{V})$ converts each column of the matrix $\bm{V}$ into a block diagonal matrix. The matrix $\bm{E}_a \in \mathbb{R}^{3n_n \times 3n_a}$ (where $n_a$ is the number of free nodes) serves as an orthogonal index matrix to isolate the free nodes $\bm{n}_a$ from all nodes $\bm{n}$, satisfying $\bm{n}_a = \bm{E}_a^T \bm{n}$. Here, $\bm{x}$ and $\bm{l}$ denote the force density and the length of members, respectively, and $\hat{\bm{v}}$ is a diagonal matrix with the components of vector $\bm{v}$ along its diagonal.
\end{theorem}

The proof is given in \cite{ma2022dynamics} using the Lagrangian method.
Given that $\bm{K}$ is a function of $\bm{n}$, Eq. (\ref{tts_equili}) is nonlinear with respect to $\bm{n}$. The process of determining the final configuration from specified loading conditions and an initial configuration can be solved by Algorithm \ref{form_finding}.

\begin{theorem}[Tensegrity Modal Information]\label{modal}
The modal information of a tensegrity structure can be determined by solving the generalized eigenvalue problem:
\begin{align}
& \bm{K}_{Taa} \bm{\varphi}=\omega^{2}\bm{M}_{aa} \bm{\varphi},
\label{eig_problem}
\end{align}
where:
\begin{align}
& \bm{M} =\frac {1} {6}(|\bm{C}|^T\hat{\bm{m}}|\bm{C}|+ \lfloor|\bm{C}|^T\hat{\bm{m}}|\bm{C}|\rfloor)\otimes \bm{{I}}_3 \\
&    \bm{K}_{Taa}  = \bm{E}_a^T \bm{K}_T \bm{E}_a, ~ \bm{M}_{aa} = \bm{E}_a^T \bm{M} \bm{E}_a,\\ \label{kt}
&    \bm{K}_T = (\bm{C}^T\hat{\bm{x}}\bm{C})\otimes \bm{I}_3 +  \bm{A}_1 \hat{\bm{E}} \hat{\bm{A}} \hat{\bm{l}}^{-3}  \bm{A}_1^T,\\  \label{A1}
&    \bm{A}_1 = (\bm{C}^T\otimes \bm{I}_3)\textit{b.d.}(\bm{N}\bm{C}^T),
\end{align}
where $\bm{E}$ and $\bm{A}$ in $\mathbb{R}^{n_e}$ denote Young's modulus and cross-sectional area of all elements, respectively. The variable $\omega$ represents the natural frequency of the tensegrity system, and $\bm{\varphi}$ is the eigenvector corresponding to the mode shape.
\end{theorem}
By linearizing the dynamics equation and omitting the external force and damping terms, one can derive the generalized eigenvalue equation, and details are provided in \cite{ma2022dynamics}.

\begin{algorithm}[ht!]
\caption{Solutions to the Tensegrity Form-Finding and Physical Properties.}
   {\bf 1)} Initialization: Given structure initial configuration $\bm{n}^0$, free nodal index matrix $\bm{E}_a$, connectivity matrix $\bm{C}$, prestress $\bm{x}_0$ in the structure members, external force $\bm{f_{ex}}$, and equilibrium computational tolerance $\epsilon$ and a positive scalar $\mu$, i.e., $\epsilon = 10^{-6}$ and $\mu = 0.1$.\\
   {\bf 2)} Input: the change of rest length $\mathrm{d}\bm{l}_{0}$ of the members. \\
    {\bf 3)} Compute rest length of the members $\bar{\bm{l}}_0 = \bm{l}_0 + \mathrm{d}\bm{l}_{0} $, the force density $\bm{x} = \hat{\bm{E}}\hat{\bm{A}}(\Bar{\bm{l}}_0^{-1}-{\bm{l}}^{-1})$ and the initial stiffness $\bm{K} | _{\bm{n}^0}$ and out of balance force: $\bm{f}_a^0 = \bm{E}_a^T(\bm{f}_{ex} - \bm{g}- \bm{K}|_{\bm{n}^0}\bm{n}^0)$. \\
\While{$||\bm{f}^i_a|| > \epsilon$ }{
Compute $\bm{K}_{Taa}$, set $\lambda \leftarrow \min\{eig(\bm{K}_{Taa})\}$. Modify the tangent stiffness matrix:\\
~~~~~~$\Tilde{\bm{K}}_{Taa} = \begin{cases}
\bm{K}_{Taa} + (\mu + |\lambda|)\bm{I},~ \lambda<0,\\ \bm{K}_{Taa} + \mu \bm{I},~ \lambda \geq 0.
\end{cases} $. \\
Compute the increment of the nodal displacement:
$\mathrm{d}\bm{n}_a^i = \Tilde{\bm{K}}_{Taa}^{-1} \bm{f}^{i-1}_a$. \\
{Compute $\delta$ from} $\begin{cases}
\underset{\delta}{\text{min}}~
V, \\ \text{s.t.}~0<\delta\leq 1.
\end{cases}. $ \\
{Line search method: $\bm{n}_a^i = \bm{n}_a^{i-1} + \delta\mathrm{d}\bm{n}_a^i$.}
\\
Compute the out of balance force: $\bm{f}_a^i = \bm{E}_a^T(\bm{f}_{ex} - \bm{g} - \bm{K} | _{\bm{n}^i}\bm{n}^i )$.
\\
$i \leftarrow i+1$.}
Output: The coordinate $\bm{n}_a$ at the final state, member force vector $\bm{t}$, and natural frequency $\bm{\omega}$ from Eq. (\ref{eig_problem}).
\label{form_finding}
\end{algorithm}

\section{Data-Based Modeling by Deep Neural Networks}\label{dnn}

In this work, we apply the sequential feedforward DNN in Keras \cite{chollet2015keras}, which consists of multiple hidden layers between the input and output layers. Each layer typically applies a nonlinear transformation to its input, allowing the network to learn complex patterns in the data. We aim to use DNN to learn tensegrity forms, which are encoded by the node coordinate information, and the physical properties of tensegrity structures, such as tension forces and frequencies. To be specific, we use $\bm{y}=(\bm{n}_a,\bm{t},\bm{\omega})$ to denote the solution estimated by DNN and assume that the DNN architecture consists of $L+1$ layers with the middle $L-1$ hidden layers. To find the forward solution, the first layer takes the change of cable rest length $\bm{l}_0$ as the input, and the last layer produces the solution $\bm{y}$ as the output. Let $z_i^{\ell}$ be the value of the $i$-th neuron at the $\ell$-th layer. Then the $j$-th neuron at the $(\ell+1)$-th layer has the value as:
\begin{align}
z_j^{\ell+1}=\sigma_{\ell+1}\left(\sum_{i=1}^{m_\ell}w_{ji}^{\ell+1}z_i^{\ell}+b^{\ell+1}\right),
\end{align}
where $\sigma_{\ell+1}$ is the activation function, $w_{ji}^{\ell+1}$ is the weight between the neurons $z_j$ and $z_i$, and $b^{\ell+1}$ is the bias term at the $(\ell+1)$-th layer. One common option for the activation function is the Rectified Linear Unit (ReLU), which sets all negative values to zero and keeps positive values unchanged. Note that the number of neurons at the $\ell$-th layer $m_{\ell}$ can be distinct across layers. Therefore, the output of the $(L+1)$-th layer from the entire network is the desired solution, i.e., $\bm{y} =(z_0^{L+1},\ldots,z_{m_{L+1}}^{L+1})$. Through the training stage, we intend to obtain the optimal weights and biases to minimize some loss function for the network, e.g., mean squared error (MSE) between the predicted outputs and the actual outputs. Then, a stochastic optimization algorithm such as Adaptive moment estimation (Adam) \cite{kingma2014adam} can be applied in the backpropagation step of DNN to obtain the optimal weights and biases. The prediction accuracy of the learned DNN model for predicting node coordinates and physical properties of specific tensegrity structures can be evaluated through the testing stage.

\section{Numerical Experiments}\label{exp}

In this section, we conduct numerical experiments on three types of tensegrity structures, including a D-bar unit, prism, and lander, to justify the effectiveness of the proposed framework in solving form-finding problems. To evaluate prediction accuracy, we use the average MSE over 20 trials to reduce the randomness of the results, i.e., each random test is repeated 20 times. Throughout the experiments, we generate data sets by applying Algorithm \ref{form_finding} with various selections of rest lengths of the cables. In particular, we focus on a uniform random selection of cable lengths. To standardize data ranges and enhance learning accuracy, we divide the forces and the frequencies by $10^3$ N and $10^6$ Hz, respectively. We generate five datasets consisting of 1,000, 2,000, 3,000, 4,000, and 5,000 data samples for all the examples. In the DNN, we follow the 80-20 train-test split, i.e., the first 80\% of the randomly shuffled dataset is allocated to the training set, with the remainder assigned to the test set.

All the numerical experiments are implemented on Python 3 in a desktop computer with Intel CPU i9-9960X RAM 64GB and GPU Dual Nvidia Quadro RTX5000 with Windows 10 Pro.
In addition, a DNN model is built in Keras \cite{chollet2015keras}, which uses the ReLU and Adam as the respective default activation function and optimizer without otherwise specified. The learning rate is fixed as 0.01, the number of units at each hidden layer is set as 64, and three hidden layers and 200 epochs for the training stage are used by default. All of these DNN configurations have empirically demonstrated the best performance in our experiments.

All the bars and strings are made of linear elastic steel with a density of 7,850 kg/m$^3$, a yield strength of 300 MPa, and a Young's modulus of 200 GPa. The bars are hollow, featuring an outer radius of 10 mm and an inner radius of 8 mm, while the strings have a radius of 2 mm.

\subsection{Tensegrity D-Bar Unit}
\begin{figure}[ht]
\centering
\includegraphics[width=.2\textwidth]{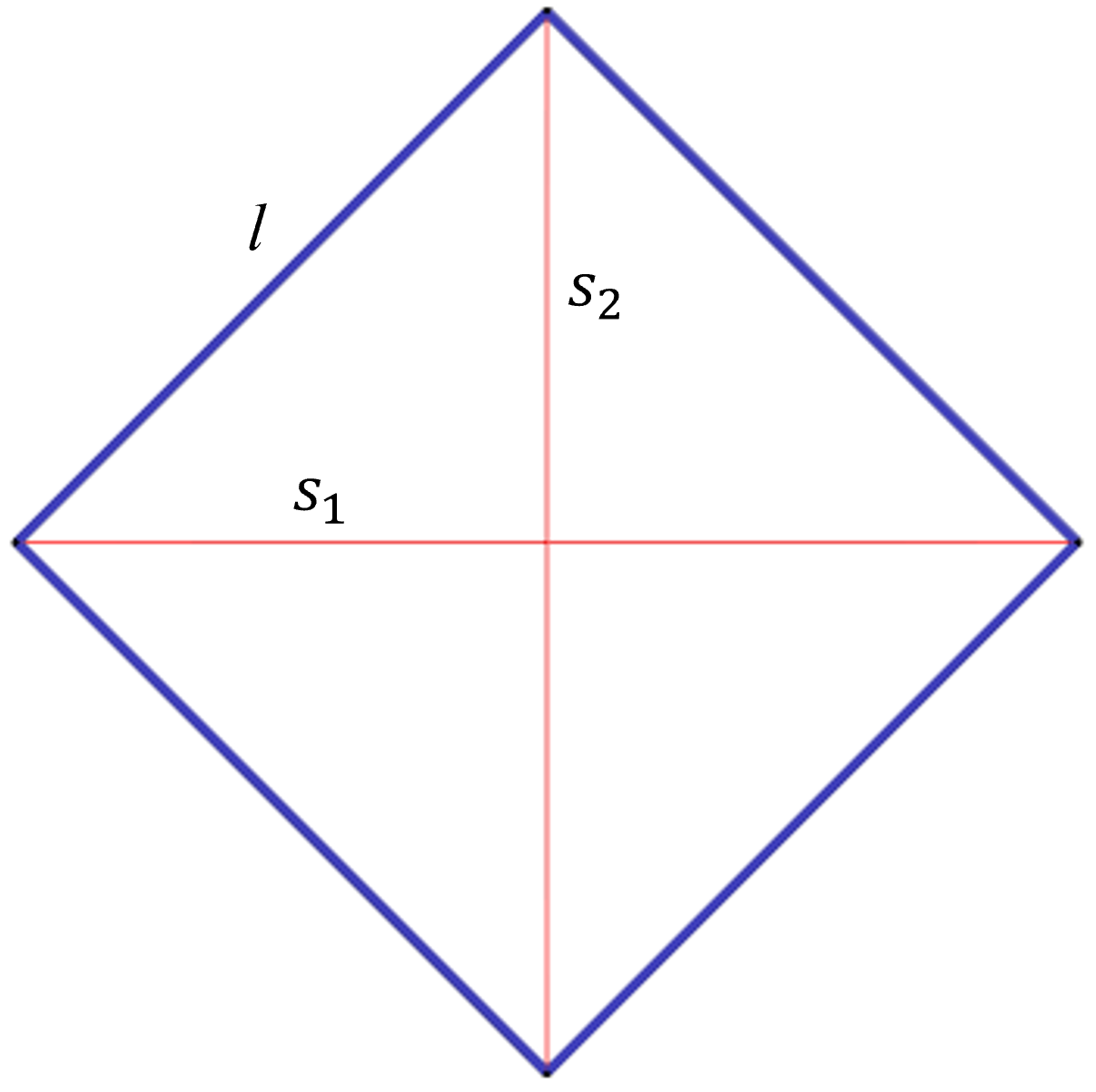}
\vspace{-10pt}
\caption{Geometric configuration of D-Bar Unit. The blue and red lines are bars and strings, respectively. There are four bars and two strings in this unit. The bar length is $l= \sqrt{2}$ m.}\label{Dbar}
\end{figure}

In the first experiment, we consider the D-bar unit tensegrity structure, as shown in Fig.~\ref{Dbar}.

\begin{figure}[ht]
\centering
\includegraphics[width=.32\textwidth]{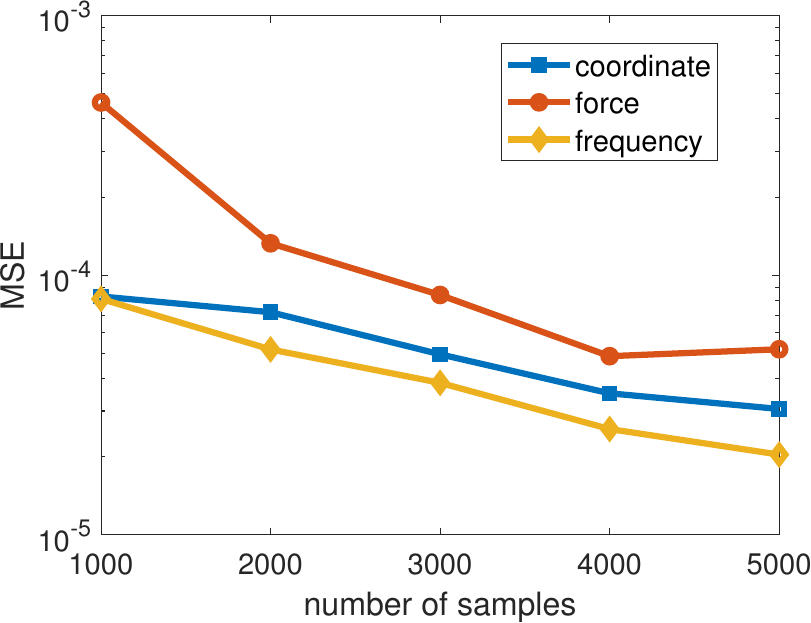}
\vspace{-10pt}
\caption{Prediction errors for the D-bar dataset.}\label{fig:dbar_mse}
\end{figure}

Each data sample has two cable lengths, and the outputs consist of reduced nodal coordinates as the inputs due to the symmetry of the D-bar (only 2 coordinates), 6 member forces, and 6 non-zero frequencies. Each cable rest length change varies uniformly randomly within $[-1,0]$ m (negative means shorten the strings). Average prediction errors in terms of average MSE for solving form-finding problems using various numbers of data samples are shown in Fig.~\ref{fig:dbar_mse}.

\begin{table}[ht]
    \caption{Runtime for the D-bar dataset.}\label{tab:dbar_time}
\centering
    \begin{tabular}{c|c|c|c|c|c}
    \hline\hline
   Sample No. & 1,000&2,000&3,000&4,000&5,000\\ \hline
   Train [s]& 9.15 &  18.65  & 31.75  & 48.78 &  94.62\\
   Test [s] &11.03 &  23.23  & 36.88 &  48.01  & 62.45\\
   \hline\hline
    \end{tabular}
\end{table}

One can observe that force errors largely dominate the overall prediction errors, decaying the fastest to 5.19 $\times 10^{-5}$, while coordinate and frequency errors tend to decay more slowly or even oscillate slightly as the number of data samples increases. The average training and testing runtimes for all the sampling cases are shown in Table~\ref{tab:dbar_time}. Overall, the growth in data size leads to improved prediction accuracy, although individual components may exhibit different decay behaviors.

\subsection{Tensegrity Prism}
\begin{figure}[ht]
\centering
\includegraphics[width=.32\textwidth]{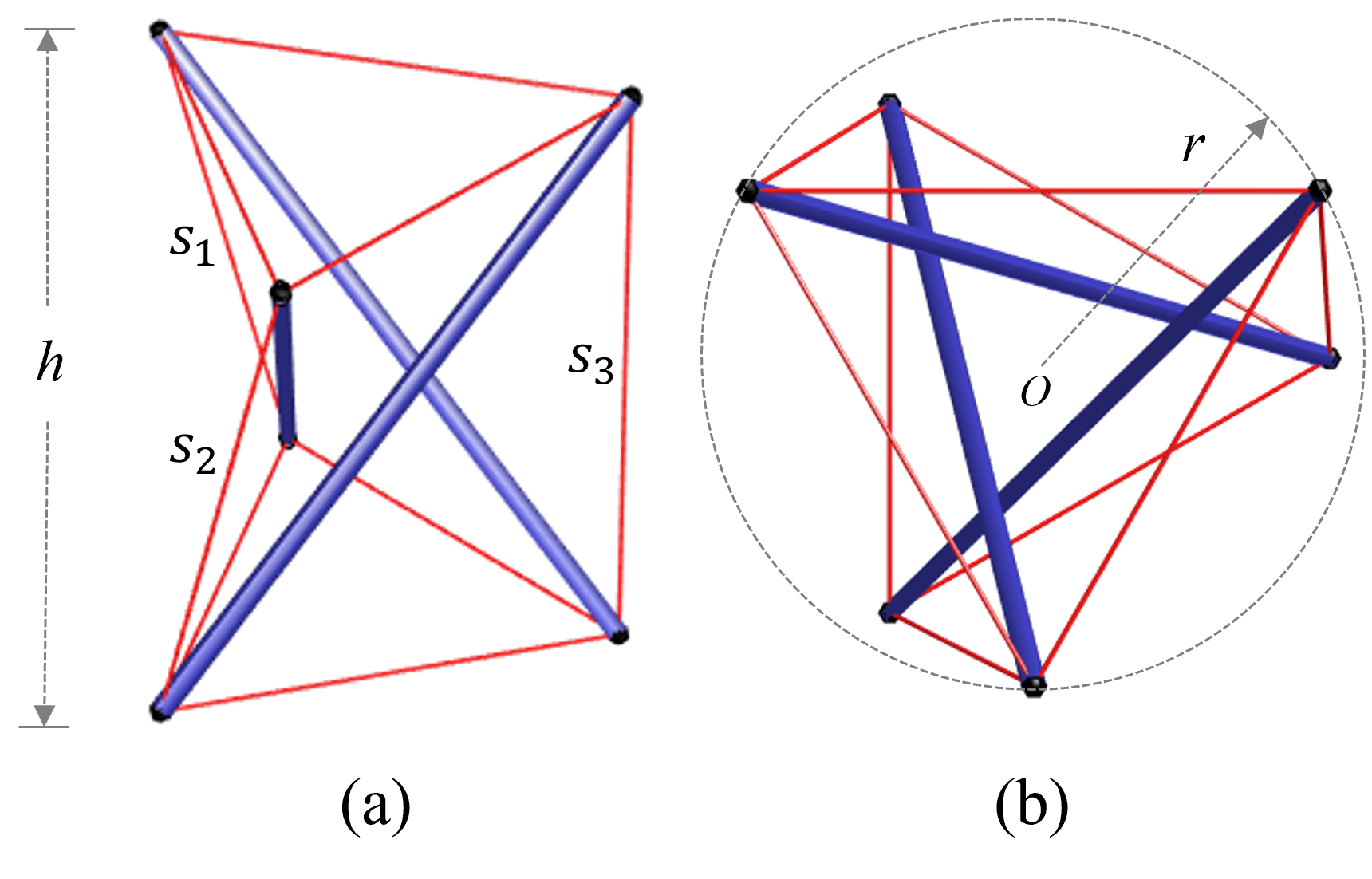}
\vspace{-12pt}
\caption{Prism geometric configuration: (a) oblique view and (b) top view. The radius and height are $r=0.25$ m and $h=0.5$ m.}\label{prism}
\end{figure}

Next, we consider the prism tensegrity structure; see Fig.~\ref{prism}.

\begin{figure}[ht]
\centering
\includegraphics[width=.32\textwidth]{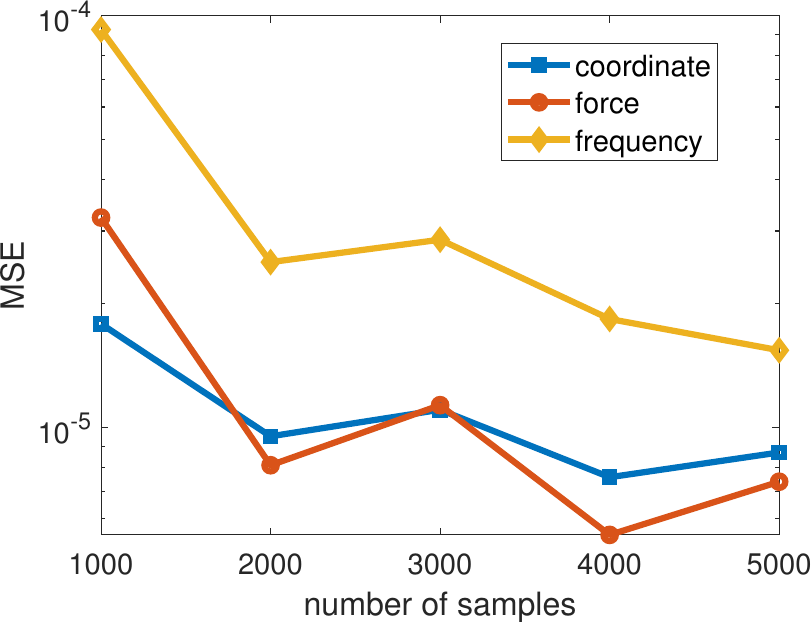}
\vspace{-10pt}
\caption{Prediction errors for the prism dataset.}\label{fig:prism_mse}
\end{figure}

Here, each sample consists of three cable lengths as inputs, eighteen nodal coordinates, twelve force values, and twelve non-zero frequency values as outputs. Similar to the D-bar datasets, we select the three vertical strings to change their rest length; the rest length of each cable changes within $[-0.15,0]$ m using a uniform random sampling scheme. Average prediction errors in terms of average MSE using various numbers of samples are plotted in Fig.~\ref{fig:prism_mse}.

\begin{table}[ht]
    \caption{Runtime for the prism dataset.}\label{tab:prism_time}
\centering
    \begin{tabular}{c|c|c|c|c|c}
    \hline\hline
   Sample No. & 1,000&2,000&3,000&4,000&5,000\\ \hline
   Train [s]& 9.23  & 17.45   &25.16  &52.17 &107.22\\
   Test [s]&11.15   &22.60  &34.53  & 50.47 &  68.39\\
   \hline\hline
    \end{tabular}
\end{table}

One can observe that frequency errors dominate the overall prediction errors, decaying the fastest to 1.54 $\times 10^{-5}$, while the decays of force and coordinate errors are very slow or even stable. The training and testing runtimes for all the sampling cases are presented in Table~\ref{tab:prism_time}. Overall, as the number of data samples increases, the overall prediction error decreases, although some components behave differently.

\subsection{Tensegrity Lander}
\begin{figure}[ht]
\centering
\includegraphics[width=.32\textwidth]{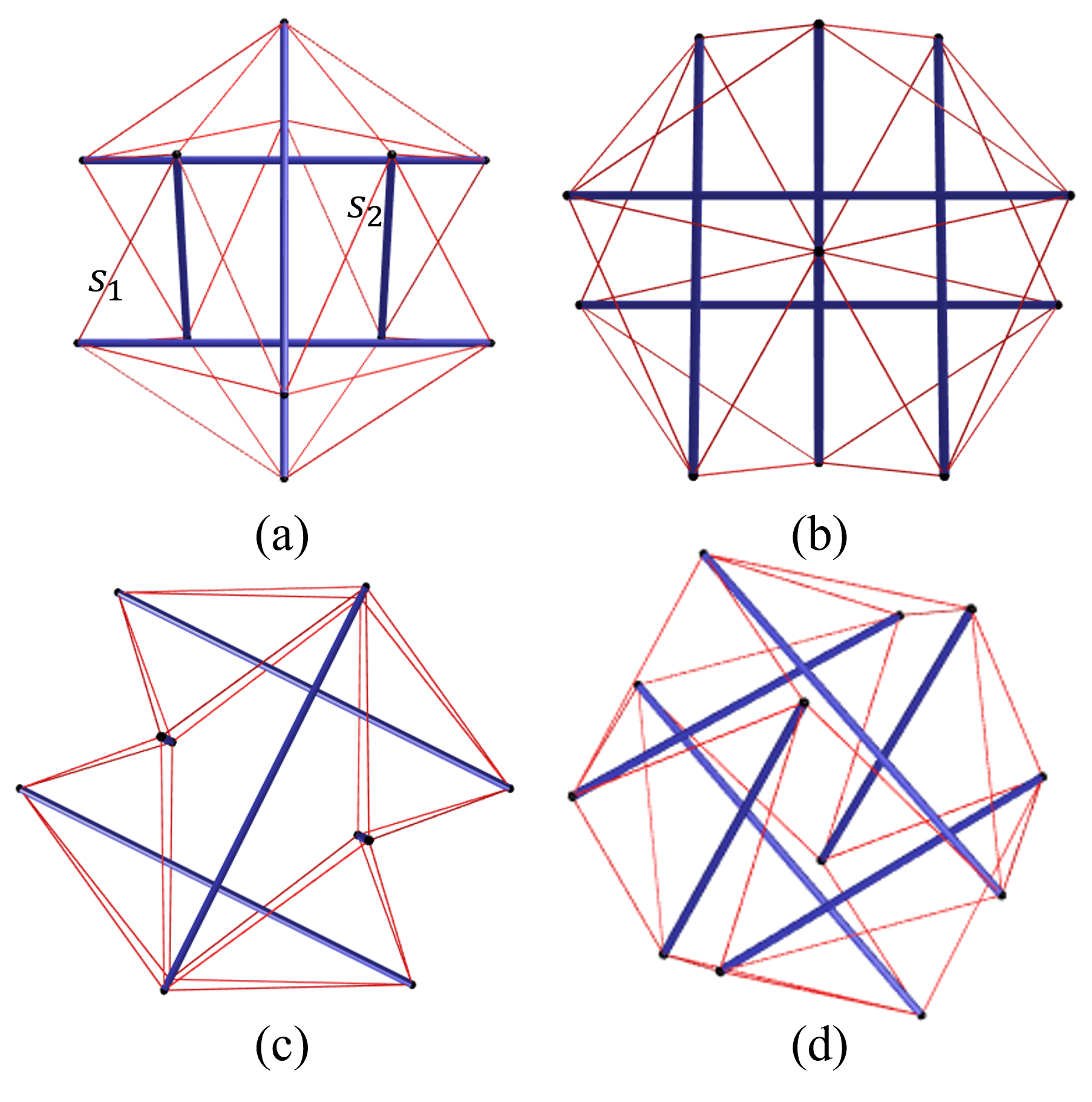}
\vspace{-10pt}
\caption{Six-bar tensegrity lander geometric configuration: (a) front view, (b) side view, (c) top view, and (d) oblique view. The bar length is 1 m. }\label{fig:lander}
\vspace{-12pt}
\end{figure}

In the third experiment, we investigate a six-bar tensegrity lander, as shown in Fig. \ref{fig:lander}.

Each data sample consists of 2 cable lengths, 36 free nodal coordinates corresponding to 12 nodes, 30 member forces, and 30 non-zero frequencies. In other words, there are two inputs and 96 outputs for the DNN. The selections of cable lengths are similar to the D-bar datasets; we select two adjacent strings to change their rest length, and the rest length of each cable changes within $[-0.3,0]$ m using a uniform random sampling scheme.

\begin{figure}[ht]
\centering
\includegraphics[width=.32\textwidth]{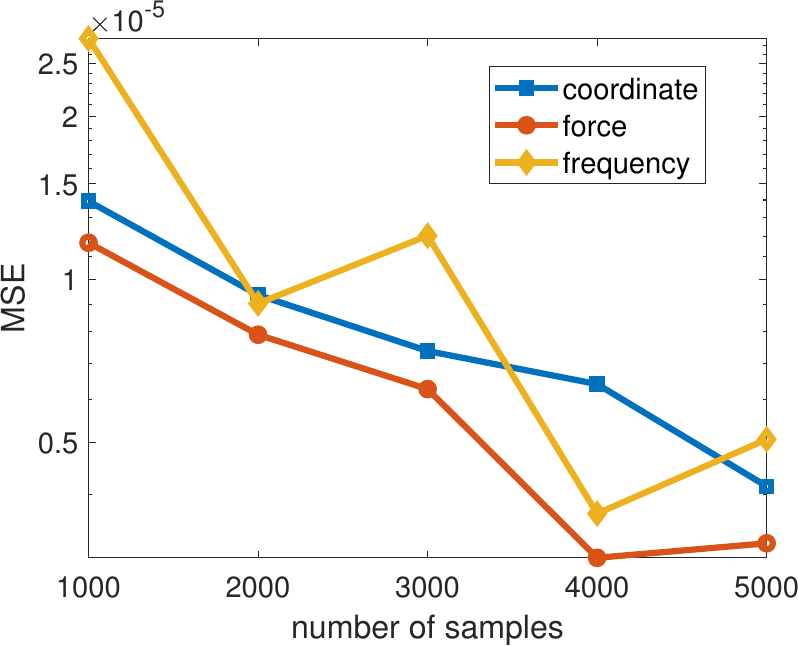}
\vspace{-10pt}
\caption{Prediction errors for the lander dataset.}\label{fig:lander_mse}
\end{figure}

\begin{table}[ht]
    \caption{Runtime for the lander dataset.}\label{tab:lander_time}
\centering
    \begin{tabular}{c|c|c|c|c|c}
    \hline\hline
   Sample No. & 1,000&2,000&3,000&4,000&5,000\\ \hline
   Train [s]& 78.84  & 154.63   &239.59  &381.19 &572.22\\
   Test [s]&13.38   &26.22 &46.50  & 63.19 &  80.33\\
   \hline\hline
    \end{tabular}
\end{table}

The average prediction errors in terms of average MSE for this dataset are shown in Fig.~\ref{fig:lander_mse}, which indicates the frequency error dominates the entire error with a similar pattern as the prism but can achieve higher accuracy (5.07 $\times 10^{-6}$ compared to 1.54 $\times 10^{-5}$). The runtimes for the training and testing are reported in Table~\ref{tab:lander_time}. As the number of data samples increases, both the training and testing times grow while the prediction accuracy improves. Compared to the D-bar and prism datasets, the lander dataset achieves a higher prediction accuracy with the same number of data samples. This difference can be attributed to the lander dataset having a much larger number of outputs compared to the other datasets.

\section{Conclusion}
\label{conc}

This paper introduces a DNN-based method to learn and predict the form-finding and physical properties of any tensegrity structures in equilibrium. Initially, we present the analytical equations governing these structures, addressing statics that include nodal coordinates and member forces, as well as natural frequencies.
We then propose to train DNN models that can predict both tensegrity forms and physical properties without explicitly solving the equilibrium equations.
To validate our approach, we examine three tensegrity structures: a D-bar, prism, and lander. The results indicate that: (1) The proposed approach achieves low output errors, demonstrated by the MSE of forces in the lander example being less than $3.26 \times 10^{-6}$ with 5,000 samples. (2) The trained DNN prediction model performs better for complex structures, as shown by the decreasing MSE from the simpler 2D D-Bar to the more complex 3D prism and lander. (3) Increasing the sample size enhances the DNN's accuracy, evidenced by the reduction in coordinate MSE in the lander from $1.39 \times 10^{-5}$ with 1,000 samples to $4.15 \times 10^{-6}$ with 5,000 samples. (4) The MSE for coordinates, forces, and frequencies are all within the same order of magnitude. This technique is versatile, applicable to various tensegrity structures, and extendable to other areas of structural physics for information identification.

\bibliographystyle{IEEEbib}

\bibliography{references}
\end{sloppypar}

\end{document}